\documentclass[letterpaper, english, reprint, aps, prl, amsmath, amssymb, floatfix]{revtex4-2}

\usepackage[T1]{fontenc}
\usepackage[utf8]{inputenc}
\usepackage{babel}
\usepackage[dvipsnames]{xcolor}
\usepackage{graphicx}
\usepackage[unicode=true,bookmarks=false,backref=false]{hyperref} 
\hypersetup{breaklinks=true,pdfborder={0 0 0},pdfborderstyle={},colorlinks=true}
\hypersetup{
    pdftitle={Winding Numbers and Topology of Aperiodic Tilings},
    pdfauthor={Yaroslav Don and Eric Akkermans},
    pdfencoding=auto
}
\hypersetup{citecolor=[rgb]{0.3,0.0,0.7}}

\usepackage{etoolbox}
\usepackage{pdfpages}
\usepackage{mathtools}
\usepackage{textcomp}
\usepackage[all]{xy}

\makeatletter

\AtBeginDocument{\let\LS@rot\@undefined}

\newcommand{\supplementary}{%
    \@for\suppage:={1,2,3,4,5,6}\do{%
        \includepdf[pages=\suppage]{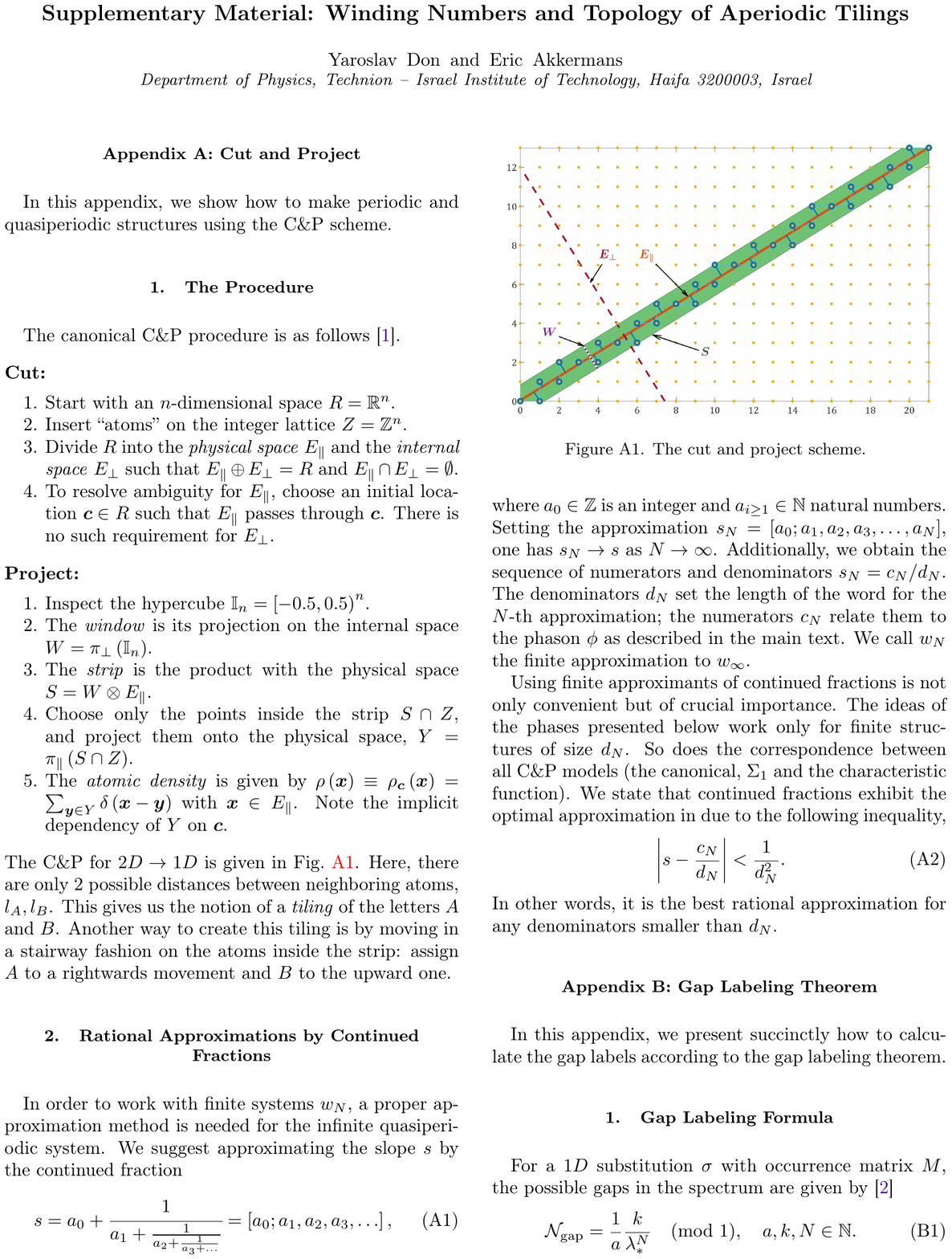}%
    }%
}

\newcommand{\appref}[1]{%
    \citep[][Sec.\ %
        \ifstrequal{#1}{app:sec:Cut-and-Project}{A1}{%
        \ifstrequal{#1}{app:sec:Continued-fractions}{A2}{%
        \ifstrequal{#1}{app:sec:Gap-labeling}{B}{%
        \ifstrequal{#1}{app:sec:Bratteli-Diagrams}{C1}{%
        \ifstrequal{#1}{app:sec:Cech-calculation}{C2}{%
        \ifstrequal{#1}{app:sec:Complexity}{C3}{%
        \ifstrequal{#1}{app:sec:Scattering-Matrix}{D2}{%
        }}}}}}}%
    ]{Supplementary}%
}

\makeatother

\def\node{*+[o][F]}    
\def\arst{\ar@{-}}     
\def\arcu{\ar@/^/}     
\def\arcd{\ar@/_/}     
\def\arcU{\ar@/^2ex/}  
\def\arcD{\ar@/_2ex/}  
\let\oldxymatrix\xymatrix
\newcommand{\smallxy}{\renewcommand{\xymatrix}{\oldxymatrix @C=1.0em @R=0.25em} }
\newcommand{\smallerxy}{\renewcommand{\xymatrix}{\oldxymatrix @C=0.5em @R=0.25em} }
\newcommand{\bigxy}{\let\xymatrix\oldxymatrix }
\smallxy

\newcommand\I{\mkern1mu {\operatorfont i}\mkern1mu}%
\newcommand\E{\operatorname{e}}%
\newcommand{\transpose}{\mathsf{T}}%
\newcommand{\IDOS}{\mathcal{N}}
\newcommand{\Smatrix}{\mathcal{S}}
\newcommand{\winding}{\mathcal{W}}
\newcommand{\NN}{\mathbb{N}}
\newcommand{\ZZ}{\mathbb{Z}}
\newcommand{\QQ}{\mathbb{Q}}
\newcommand{\RR}{\mathbb{R}}
\newcommand{\CH}{\check{H}^{1}}

\DeclareMathOperator\im{Im}%
\DeclareMathOperator\diag{diag}%
\DeclareMathOperator\tr{Tr}%
\newcommand{\vcen}[1]{\vcenter{\vbox{#1}}}%

\allowdisplaybreaks

\begin{document}

\title{Winding Numbers and Topology of Aperiodic Tilings}

\author{Yaroslav Don}
\author{Eric Akkermans}
\email{eric@physics.technion.ac.il}
\affiliation{Department of Physics, Technion -- Israel Institute of Technology, Haifa 3200003, Israel}

\date{\today}

\begin{abstract}
We show that diffraction features of $1D$ quasicrystals can be retrieved from a single topological quantity, the \v{C}ech cohomology group, $\CH \cong\ZZ^2$, which encodes all relevant combinatorial information of tilings. We present a constructive way to calculate $\CH$ for a large variety of aperiodic tilings.
By means of two winding numbers,
we compare the diffraction features contained in $\CH$ to the gap labeling theorem, another topological tool used to label spectral gaps in the integrated density of states. 
In the light of this topological description, we discuss similarities and differences between families of aperiodic tilings, and the resilience of topological features against perturbations.   
\end{abstract}

\maketitle


Aperiodic tilings are structures obtained from the spatial arrangement of motives according to a set of deterministic rules~\citep{Senechal_Book_1996,Mozes_JAM_1989}. They constitute a rich playground to investigate a wealth of new ideas and features of physical systems in different contexts, such as condensed matter, statistical mechanics, dynamical systems and new materials. This ubiquity is partly due to the existence of a large set of tiling families including periodic, nonperiodic (e.g.\ Wang tiles), quasiperiodic, and asymptotically periodic tilings. 

A celebrated family of aperiodic tilings are quasicrystals~\citep{Levine_1984,Shechtman_1984}. 
Despite their lack of periodicity, they exhibit  Bragg peaks. Spectral characteristics of propagating waves (acoustic, optical, matter) in quasicrystals reveal a highly lacunar fractal energy spectrum, with an infinite set of energy gaps~\citep{Luck_PRB_1989,Damanik_CMP_2008,Tanese_2014,Sutherland_PRB_1986}. The gap labeling theorem (GLT)~\citep{Bellissard_RMP_1991,Bellissard_Review_1992} provides a framework for the description of these gaps.
Where Bloch theorem allows to label energy eigenstates with a quasi-momentum and to identify topological numbers expressed in terms of a Berry curvature~\citep{TKNN_1982,Berry_1984,Thouless_1983,Xiao_2010}, 
the GLT assigns a set of integers to each gap in the spectrum of aperiodic tilings. These integers can be given a topological meaning akin in nature to Chern numbers or alike, but different in many respects and not expressible in terms of a curvature~\citep{Bellissard_RMP_1991,Kunz_1986}. Topological features of aperiodic tilings are still largely uncharted despite extensive interest, and situations where they can be directly measured in an experiment remain an exception~\citep{Poddubny_Ivchenko_PE_2010,Kraus_2012a,Kraus_2012b, Bandres_2016,Baboux_2017,Dareau_PRL_2017}. A reason for this state of affairs is that  topological features of tilings are often established in the limit of infinite systems, whereas finite size measurable aperiodic samples cannot be unequivocally discriminated from periodic tilings of appropriate unit cells. As a consequence, topological properties  of aperiodic tilings are often dimmed and considered irrelevant~\citep{Dana_2014,Poshakinskiy_2015,Madsen_2013}.

Our purpose is to present a systematic and easily implemented  description of topological aspects of aperiodic tilings in terms of (integer) winding numbers of two phases respectively associated to structural and spectral features. This description holds for finite size tilings in any dimension. For quasicrystals, these two phases are equivalent, thus leading to an extension of Bloch theorem beyond periodic systems~\citep{ADRS_2021_Bloch}. For other aperiodic tilings, this equivalence does not hold. Finally, our approach offers a simple description of GLT and related topological aspects which, despite their importance, are often difficult to decipher. In addition, it will help unifying disparate accounts of structural and spectral properties of tilings scattered in different parts of an abundant literature on that subject.


To convey essential ideas of our description and to define the two phases and their winding numbers, we first consider $1D$ tilings built out of a two-letter $\left\{ A,B\right\} $ alphabet, each representing a tile of respective length $\left\{ l_{A},l_{B}\right\} $ with an atom inserted  on the boundary between  adjacent tiles. 
The atomic density is $\rho\left(x\right)=\sum_{i}\delta \left(x-x_{i}\right)$, where
$x_{i}=\smash{\sum_{j=1}^{i}l_{j}}$ is the atom location. The Fourier transform $\hat{\rho}\left(k\right)$ of $\rho$ is a complex valued function whose modulus determines the structure
factor $S\left(k\right)=\left|\hat{\rho}\left(k\right)\right|^{2}$ of the tiling namely its diffraction spectrum. We denote  $\Theta_d (k) = \arg \hat{\rho}\left(k\right)$ the phase of $\hat{\rho}\left(k\right)$. It is the first relevant phase we consider and it accounts for structural data of tilings.

The band structure of excitations (e.g.\ electronic, electromagnetic, acoustic or mechanical waves) propagating in a tiling is modeled either by a tight-binding model with particles hopping from tile to tile or by a continuous wave equation (e.g.\ Schr\"odinger or Helmholtz). The quantum/wave  mechanical description involves a certain self-adjoint operator in the space of square-summable functions in the set of tiles.  We are interested in its energy/frequency 
spectrum, a well documented problem in condensed matter  literature \citep{Akkermans_Review_2014}. The counting function $\IDOS(E)$, or integrated density of states (IDOS), is the fraction of eigenenergies  smaller than a given value $E$. For large enough system size, $\IDOS(E)$ is independent of the choice of boundary conditions and it is a well defined and continuous function of energy. 
A convenient description of spectral properties is provided by the scattering matrix formalism where, in $1D$, a $2 \times 2$ unitary operator $\Smatrix\left(E\right)$ maps ingoing onto outgoing waves scattered by a finite size tiling. Diagonalising  $\Smatrix\left(E\right)$  leads to two independent phases: the scattering phase shift $\delta (E) =  \im\log\det\Smatrix\left(E\right) $,  allows to determine the counting function, $\IDOS(E) = \delta(E) / \pi$~\citep{Akkermans_Review_2014} (see Supplementary Material (SM)~\appref{app:sec:Scattering-Matrix}); and the chiral  phase~\citep{Levy_arXiv_2015,Levy_EPJ_2017}, 
\begin{equation}
    \Theta_s \left(E \right)=\im\tr\left[\sigma_{z}\log\Smatrix\left(E \right)\right],
    \label{eq:thetaS}
\end{equation}
where $\sigma_{z}=\diag\begin{pmatrix}1 & -1\end{pmatrix}$ is the Pauli
matrix, displays topological features. 
To unveil them, we consider the (integer) winding numbers $\winding_{\phi}\left[\Theta_d\right]$ and $\winding_{\phi}\left[\Theta_s\right]$ w.r.t.\ an angular variable $\phi$ (defined below) of the two phases $\Theta_d$ and $\Theta_s$ and show that they 
result from the existence of a topological invariant of the tiling, its Čech cohomology group $\CH$, a quantity we define and compute systematically. 
To that purpose, we start with the case of $1D$ quasicrystals. Rediscovering well established results using a new approach should not hinder that it is applicable far more generally  
as we shall see.

A quasicrystal in dimension $D$ is modeled as a section of a periodic lattice (crystal) in a  $n$-dimensional superspace $\RR^n$, with $n > D$. We have the decomposition $\RR^n = E^\parallel \oplus E^\perp$, where $E^\parallel$ is the $D$-dimensional physical space in which the structure is embedded, whereas $E^\perp$ is an $(n-D)$-dimensional internal space. This setting is implemented for the Cut \& Project algorithm (hereafter, C$\&$P)~\citep{Senechal_Book_1996,Luck_Review_1994,DuneauKatz_1985,KatzDuneau_1986}. For a $1D$-quasiperiodic tiling, the superspace is the square lattice $\ZZ^2$ and the physical space is the line $E^\parallel$ making a tilt angle $\theta$ with the horizontal axis. An irrational slope $s = 1/ (1 +\cot{ \theta})$ leads to a quasiperiodic tiling, while an irreducible rational $s = p/q$, corresponds to a periodic structure with $q$ atoms in a cell. 
A celebrated example is the Fibonacci quasicrystal obtained for the irrational slope $s = \tau^{-1} = 2/(1+\sqrt{5})$.
The location of the cut-line in the $E^{\perp}$ direction (see SM~\appref{app:sec:Cut-and-Project}) is an additional degree of freedom $\phi\in\left[0,2\pi\right]$ called phason.

\begin{figure}
\centering{}
\includegraphics[width=0.8\columnwidth]{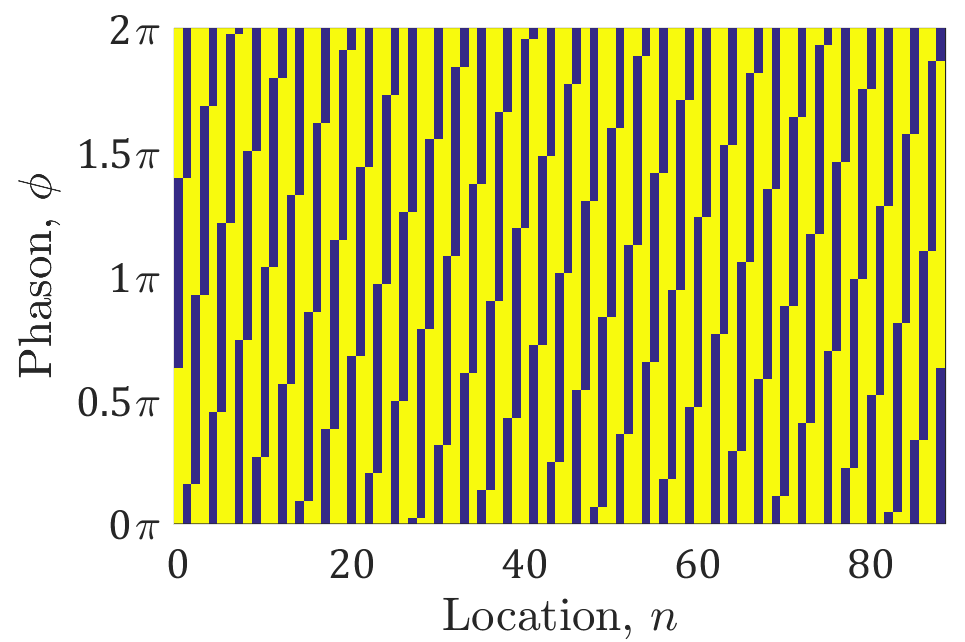}
\caption{\label{fig:Sigma1}%
    The characteristic function $\Sigma_{1}$ for the
    Fibonacci sequence with $d_{N}=89$. The $A\left(B\right)$ tiles
    with $\Sigma_{1}=+1\left(-1\right)$ are drawn in yellow (blue). Each
    row represents a word of length $d_{N}$.
}
\end{figure}
\begin{figure}
\centering{}
\includegraphics[width=1\columnwidth]{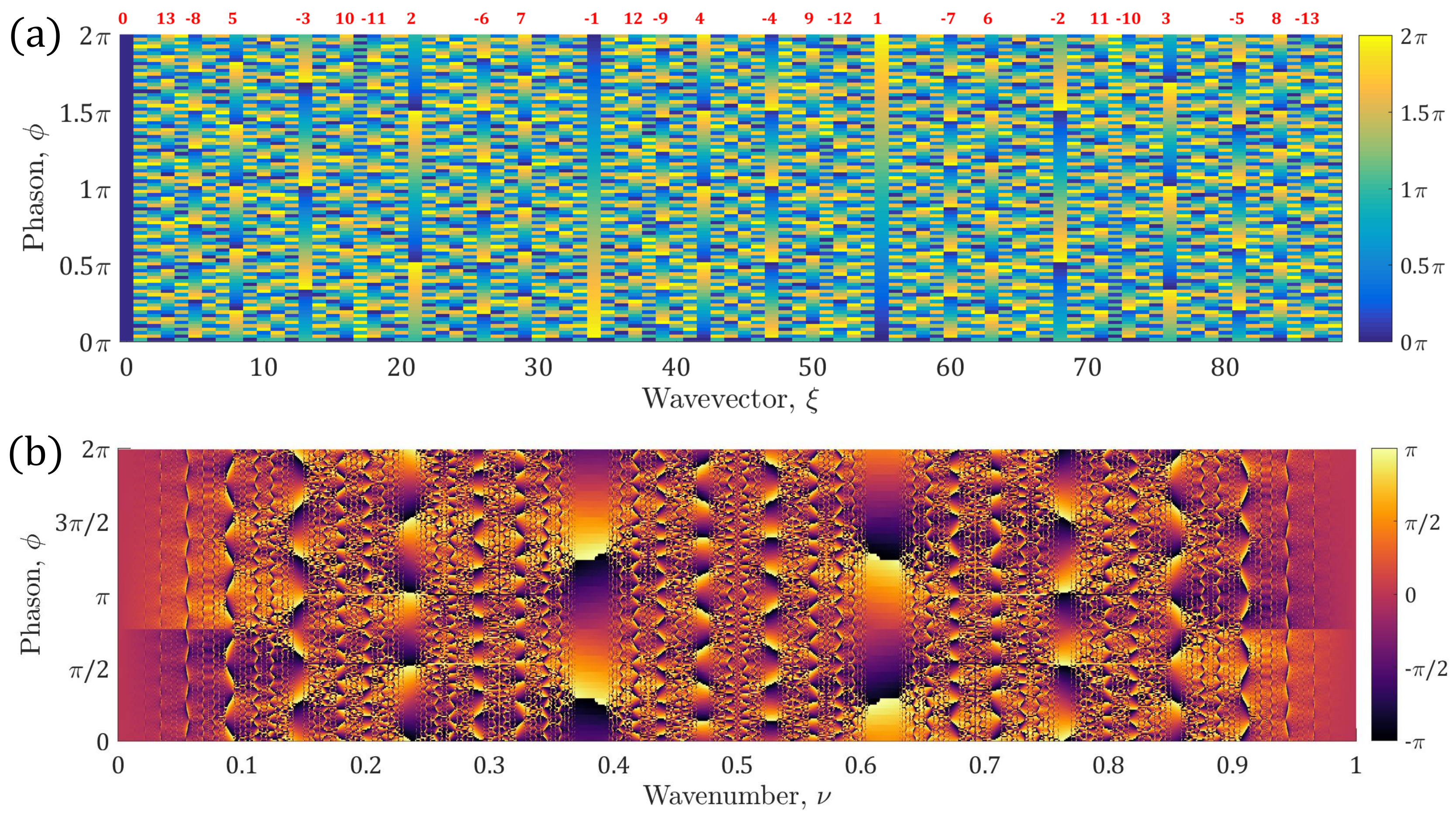}
\caption{\label{fig:topological-phases}%
    Topological phases of the Fibonacci tiling with slope $s=\left(\protect\smash{\sqrt{5}}-1\right)/2$.
    (a) The structural phase $\Theta_d\left(\xi,\phi\right)$ with $d_{N}=89$ sites. Windings are indicated by the red numbers above.
    (b) The chiral phase $\Theta_s\left(\nu,\phi\right)$ with $d_{N}=233$ sites using $\left(n_{A},n_{B}\right)=\left(1,1.15\right)$ and $\left(l_{A},l_{B}\right)=\left(1.15,1\right)$.
    The wavenumber $\nu$ is the normalized wavevector $k$. 
}
\end{figure}

To describe finite size tilings, it is convenient to 
expand the slope $s$ as a continued fraction
$c_{N}/d_{N}\xrightarrow{\smash{N\to\infty}}s$ (see SM~\appref{app:sec:Continued-fractions}) so that the phason becomes discrete, $\phi\mapsto\phi_{\ell}=2\pi\ell/d_{N}$
with $\ell=0\ldots d_{N}-1$.
The number of distinct words, i.e.\ of finite tilings, of size $L$ is 
$L+1$~\citep{Julien_2009,Robinson_Review_2004}, and for the $N$-th approximation ($L=d_{N}$), there are $d_{N}$ words, which are mutual cyclic permutations~\citep{Don_Thesis_2021}. 
Consider a representative word $w_{0}$ from the mutually-cyclic ones, and define the $d_{N}\times d_{N}$ characteristic matrix
\begin{equation}
    \Sigma_{1}\left(n,\ell\right)=\mathcal{T}^{m\left(\ell\right)}\left[w_{0}\left(n\right)\right],\label{eq:Sigma1-def}
\end{equation}
with $m\left(\ell\right)=\ell\,c_{N}^{-1}\pmod{d_{N}}$, and $\mathcal{T}\left[w_{0}\left(n\right)\right]=w_{0}\left(n+1\right)$
the translation operator with periodic boundary conditions in both
$n$ and $\ell$ directions. Thus, all the rows of $\Sigma_{1}$ are
cyclic permutations of each other. Hence, $\Sigma_{1}$ is a
torus (see Fig.~\ref{fig:Sigma1}), whose discrete Fourier transform about $n$ is 
\begin{equation}
    G_{N}\left(\xi,\ell\right)=\sum_{n=0}^{d_{N}-1}\omega^{-\xi n}\,\Sigma_{1}\left(n,\ell\right)=\omega^{m\left(\ell\right)\xi}\,\varsigma_{0}\left(\xi\right),
\label{eq:Fourier-amplitude}
\end{equation}
where $\omega=\exp\left(2\pi\I/d_{N}\right)$, and $\varsigma_{0}\left(\xi\right)$
is the Fourier transform of $w_{0}\left(n\right)$. The structure
factor $S_{N}\left(\xi,\phi\right)=\left|\varsigma_{0}\left(\xi\right)\right|^{2}/d_N$
is $\phi$-independent. However, the \emph{structural}
phase $\Theta_{d}\left(\xi,\ell\right)= \arg G_{N}\left(\xi,\ell\right)$
reads
\begin{equation}
    \Theta_{d}\left(\xi,\ell\right)=\phi_{\ell}\,\xi/c_{N}+\alpha_{0}\left(\xi\right)\pmod{2\pi}.
    \label{thetad}
\end{equation}
where $\alpha_{0}\left(\xi\right)=\arg\varsigma_{0}\left(\xi\right)$
is $\phi$-independent. Thus, for any diffraction (discrete Bragg) 
peak $\xi_{p,q}=qc_{N}$, the winding number is 
$\winding_{\phi}\left[\Theta_{d}\left(\xi_{p,q},\phi\right)\right]=q$, 
as displayed in Fig.~\hyperref[fig:topological-phases]{\ref*{fig:topological-phases}a}.

We now consider spectral features of tilings encoded in the $\phi$-dependent scattering matrix $\Smatrix\left(E,\phi\right)$ and the phase $\Theta_s \left(E,\phi \right)$ from \eqref{eq:thetaS}.
Its winding inside a gap is 
$\winding_{\phi}\left[\Theta_s\left(E_{g},\phi\right)\right]=2q$~\citep{Baboux_2017},
as displayed numerically in Fig.~\hyperref[fig:topological-phases]{\ref*{fig:topological-phases}b}. 
The winding numbers of the two phases, structural and spectral, thus fulfill~\citep{Don_Thesis_2021}
\begin{equation}
    2\winding_{\phi}\left[\Theta_d\right]=\winding_{\phi}\left[\Theta_s\right] = 2 q .
    \label{eq:windings}
\end{equation}
The integer $q$ also appears in the structure factor and the counting function $\IDOS\left(E_{g}\right)$ in the gaps~\citep{Bellissard_Review_1992}, since both the positions $k_{b}/k_{0}$ of the normalized and infinite countable set of Bragg peaks~\citep{Senechal_Book_1996} and the gap locations in the energy spectrum are related to the irrational slope $s$ by
\begin{equation}
    k_{b}/k_{0}=p+qs=\IDOS\left(E_{g}\right),\quad p,q\in\ZZ . \label{eq:Bloch-relation}
\end{equation}


This one-to-one correspondence between Bragg peaks and spectral gap locations is not coincidental, it rather reflects the equivalence between structural and spectral data of C\&P quasiperiodic $1D$ tilings. 
The meaning of $q$ as a common winding number in \eqref{eq:windings} constitutes one important result of this letter and an obvious appeal to topology.


Note that \eqref{eq:windings} does not yet provide a clear connection to topological quantities nor a systematic procedure to calculate them akin to Chern numbers and Berry curvature for periodic structures. Results similar to \eqref{eq:windings} have been obtained for $1D$ continuous wave equations in a quasi periodic potential 
\citep{Johnson_1982,Johnson_1986},
and for tight-binding operators \citep{Delyon_1983} and have been subsequently extended to more general tilings by Bellissard \citep{Bellissard_RMP_1991} and Luck \citep{Luck_PRB_1989}.  A topological interpretation of the gap labeling as a winding number has been underlined. Yet, that interpretation used the rotation number, which is essentially equivalent to the total phase shift $\delta(E)$ -- a quantity that is distinct from the winding of $\Theta_s$ w.r.t.\ the phason $\phi$. 

We now show that both diffraction and spectral features are fully characterised by a topological invariant systematically computable and directly related to the winding numbers in \eqref{eq:windings}. To  illustrate these ideas, 
we introduce the notion of substitution to generate structural data of $1D$ C\&P aperiodic tilings~\citep{Luck_JPA_1993,Luck_Review_1994,Senechal_Book_1996,Baake_Grimm_Book_2018}.

For a two-letter alphabet $\{A,B \}$, a substitution rule is defined by its action $\sigma$ on a word $w= l_1 l_2 \dots l_k$, where $l_i = A\text{ or }B$, by the concatenation $\sigma\left(w\right) = \sigma\left(l_1\right) \sigma\left(l_2\right) \dots \sigma\left(l_k\right)$. A primitive occurrence matrix, 
$ M= \smash{
\big(\begin{smallmatrix}\alpha & \beta\\
\gamma & \delta
\end{smallmatrix}\big)}$
defined by $\sigma (A) = A^\alpha B^\beta$ and $\sigma (B) = A^\gamma B^\delta$, is associated to $\sigma$.
Its largest eigenvalue $\lambda_*$ is larger than $1$ (Frobenius-Perron theorem).
Its left eigenvector $\boldsymbol{v}_{*}=(\varrho_A,\varrho_B)$ coincides with the corresponding letter densities in the infinite tiling.
For the Fibonacci substitution, $ M = \begin{psmallmatrix}1 & 1 \\
1 & 0
\end{psmallmatrix}$%
, $\lambda_{*} = \tau = \frac{1 + \smash{\sqrt{5}}}{2}$, $(\varrho_A,\varrho_B)=(\tau^{-1},\tau^{-2})$. 
The C\&P and substitution algorithms are not equivalent, e.g.\ no substitution is associated to the C\&P slope $s = 1/\pi$ since $\pi$ is a transcendental and not an algebraic irrational. Inversely, the substitution $M = 
\begin{psmallmatrix}1 & 2\\
1 & 0
\end{psmallmatrix}$ has no C\&P counterpart. But if  $\lambda_{*}$ is an algebraic rational, 
it may correspond to a C\&P tiling of slope $s=\varrho_B$, a situation that we now consider.

Denote $w_\infty$ the infinite tiling/word and consider  all $n$-letter words (also known as supertiles) $\Gamma_{n}=\left\{ w\in w_{\infty}:\left|w\right|=n\right\} $. Next, define the
shift $\gamma_{n}:\Gamma_{n}\to\Gamma_{n}$ by
$\gamma_{n}\left(w_{i}\right)=w_{j}$ if $w_{j}$ follows $w_{i}$ in $w_{\infty}$.
We define Bratteli diagrams by $G_{n}=\left(\Gamma_{n},\gamma_{n}\right)$. For example, for the Fibonacci tiling, 
\begin{subequations}
\begin{align}
    \Gamma_{1}^{\text{Fib}} & =\left\{ A,B\right\} \\
    \Gamma_{2}^{\text{Fib}} & =\left\{ AA,AB,BA\right\} \\
    \Gamma_{3}^{\text{Fib}} & =\left\{ AAB,ABA,BAA,BAB\right\} 
\end{align}
\end{subequations}
so that 
\begin{subequations}
\label{eq:Bratteli-diagrams}
\begin{align}
    G_{1}^{\text{Fib}} & =\xymatrix{\node{A}\arcu[rr]^{AB}\ar@(ul,dl)[]_{AA} &  & \node{B}\arcu[ll]^{BA}}
\\
    G_{2}^{\text{Fib}} & =\xymatrix{ & \node{AA}\arcd[ddl]_{AAB}\\
    \\
    \node{AB}\arcu[rr]^{ABA} &  & \node{BA}\arcd[uul]_{BAA}\arcu[ll]^{BAB} }
\end{align}
\end{subequations}
Each node of $G_n$ is a supertile element of $\Gamma_n$, and each edge is a supertile element of $\Gamma_{n+1}$, so that at each level $n$, the (oriented) Bratelli graph $G_n$ indicates how successive supertiles are encountered while moving along the infinite tiling $w_\infty$. 

We are interested in the number $\beta_n$ of independent closed cycles of the oriented graph $G_n$, i.e.\ in the number of independent ways to move along the tiling either finite or infinite. It is clear that $\beta_1 = \beta_2 =2$. For instance, using $\Gamma_1$ tiles, the graph $G_1$ shows that it exists two independent ways to move along a Fibonacci tiling, so that any finite length sequence can be decomposed into a linear combination of two closed cycles: ${A}\to {A}$ and ${A}\to {B}\to {A}$. Should we decide to explore the tiling using supertiles $\Gamma_2$, then again any evolution along the tiling involves a linear combination of two closed cycles: ${AB}\to {BA}\to {AB}$ and ${AA}\to {AB}\to {BA} \to {AA}$. This decomposition into linear combination of independent closed cycles of supertiles $\Gamma_n$ with integer coefficients has a group structure isomorphic to $\ZZ^{\beta_{n}}$, known as the cohomology  $H^{n}\left(G_{n},\ZZ\right)$ (see SM~\appref{app:sec:Bratteli-Diagrams}). The limiting process $n \to \infty$, if it exists, defines the Čech cohomology group $\CH$, a topological invariant of the tiling. For $1D$ C\&P quasicrystals, $\beta_n = 2$ for all $n$ (see SM~\appref{app:sec:Complexity}), namely the evolution along any sequence involves a linear combination of two cycles independently of the chosen supertile.

To systematically compute $\CH$ for aperiodic tilings, the knowledge of the application of the inflation rule on Bratteli diagrams is needed.
Consider a substitution $\sigma$ and the Bratteli diagram $G_2$ as in \eqref{eq:Bratteli-diagrams}. Next, apply $\sigma$ on $G_2$ to generate the  diagram (for the Fibonacci case)
\begin{equation}
    \smallerxy
    \sigma\left(G_2^{\text{Fib}}\right) = 
    \vcen{\xymatrix{ 
        &  & \node{AB}\arcd[dl]_{ABA}\\
        & \node{BA}\arcd[ddl]_{BAB} &  & \quad\\
        &  & \node{BA}\arcu[rrd]_{BAA}\\
        \node{AB}\arcu[rru]_{ABA} &  &  &  & \node{AA}\arcD[uuull]_{AAB}\arcu[llll]^{AAB}
    } } 
    \smallxy
\end{equation}
Identifying the nodes and edges between $G_2^{\text{Fib}}$ and $\sigma\left(G_2^{\text{Fib}}\right)$ (see SM~\appref{app:sec:Cech-calculation}), we infer the inflation matrices
\begin{equation}
    A_{1}^{\transpose}=
    \begin{psmallmatrix}
        0 & 1 & 0 & 1\\
        0 & 1 & 1 & 0\\
        1 & 0 & 0 & 0\\
        1 & 0 & 0 & 0
    \end{psmallmatrix},
    \quad A_{0}^{\transpose}=
    \begin{psmallmatrix}
        0 & 1 & 0\\
        0 & 1 & 0\\
        1 & 0 & 0
    \end{psmallmatrix}.
\end{equation}
Defining the $\zeta$-function~\citep{Anderson_Putnam_ETDS_1998},
\begin{equation}
    \zeta\left(z\right)=\frac{\det\left(I-zA_{0}^{\transpose}\right)}{\det\left(I-zA_{1}^{\transpose}\right)}=\frac{p_{0}\left(z\right)}{p_{1}\left(z\right)}\, ,
\label{eq:zeta}
\end{equation}
the Čech cohomology is then retrieved from the decomposition of the polynomials $p_0(z)$ and $p_1(z)$ into their irreducible components over the integers, and given by the direct sum of $\ZZ$ adjoined by their leading coefficients (see SM~\appref{app:sec:Cech-calculation}). 
For Fibonacci, $p_0(z)=1-z$ and $p_1(z)=1-z-z^2$ inferring 
$\CH\cong\ZZ^2$ -- independently of the irrational slope $s$. In C\&P quasicrystals, $\CH$ corresponds to the incongruent dimensions of $\RR^n$ with respect to $E^\parallel$. In $1D$ tilings originated in $\RR^2$, $\CH\cong\ZZ^2$ if $s$ is irrational, and $\CH\cong\ZZ$ if $s\in\QQ$.

We now show that winding numbers  $\winding_{\phi}\left[\Theta_d\right]$ in \eqref{eq:windings} and Bragg peaks at locations \eqref{eq:Bloch-relation}, are direct consequences of the topological property, $\CH \cong\ZZ^2$.


For irrational slopes $s$, there are $2$ cycles, which for each $N$ and $s_N=c_N/d_N$ approximant, consist of a long $d_{N}$-cycle made of $d_{N}$ cyclically permuted $d_{N}$-supertiles and a short cycle of length $d_{N-1}$. 
Since the approximant $s_N$ is rational, only the long cycle remains for $n\ge d_N-1$.
Cyclically permuting a given supertile $t_{N}$ (of length $d_N$) by $c_{N}^{-1}\pmod{d_{N}}$ tiles results in the same supertile up to a single pair switch (see Fig.~\ref{fig:Sigma1}). This is the winding property expressed in \eqref{thetad} and \eqref{eq:windings}.

To show the existence and locations of Bragg peaks, consider 
for each $N$ and $s_N=c_N/d_N$ approximant,   
the structure factor $ S_N(\xi)=|G_N(\xi)|^2/d_N$ where $\xi=k d_N$ and $G_N\left(\xi\right)$ is the discrete Fourier transform \eqref{eq:Fourier-amplitude} of supertiles $t_N$ of length $d_N$.
It has positive contributions for each $\xi_l = l+d_N p$, thus producing diffraction peaks at
$S_{N}(\xi)=\sum_{p\in\ZZ}\sum_{l=0}^{d_{N}-1}a_{l}\,\delta_{\xi,l+d_{N}p}$.
Furthermore, using the winding property above, the diffraction of \emph{permuted} supertiles reads,
$\tilde{S}_{N}(\xi) \simeq d_{N}^{-1} \E^{2\pi\I c_{N}^{-1}\xi}|G_N(\xi)|^2 = \E^{2\pi\I c_{N}^{-1}\xi} S_{N}(\xi) $.
Since $S_N$ and $\tilde{S}_N$ must match, then $\E^{2\pi\I c_{N}^{-1}\xi}=1$ implying $\xi_{q}=c_{N}q$. Hence,
\begin{equation}
    S_{N}\left(\xi\right)=\sum_{p\in\ZZ}\sum_{q=0}^{d_{N}-1}a_{q}\,\delta_{\xi,c_{N}q+d_{N}p} \, .
    \label{eq:periodic-cnp-diff}
\end{equation}
Overall, there are two copies of $S_{N}(k)$ at each order $N$.
Changing to $k=\xi/d_N$ in \eqref{eq:periodic-cnp-diff} and taking the limit $N\to\infty$ gives~\citep{Don_Thesis_2021}
\begin{equation}
    S\left(k\right)=\sum_{p\in\ZZ}\sum_{q\in\ZZ} \bar{a}_{q}\,\delta\left(k-\left(p+sq\right)\right),
\label{eq:CnP-SF}
\end{equation}
as announced in \eqref{eq:Bloch-relation}. 

We have thus proven that for $1D$ C\&P tilings, the topological invariant $\CH \cong\ZZ^2$ and the associated winding numbers \eqref{eq:windings}, provide the set of integers needed to describe the structure factor \eqref{eq:CnP-SF}. These results have been summarized~\citep{ADRS_2021_Bloch} using a trace map, denoted $\tau_* ^{d} \left(\smash{\CH}\right)$ between $\CH \cong\ZZ^2$ 
and the group $\ZZ + s \ZZ$, namely  
$\tau_* ^{d} \left(\smash{\CH} \cong \ZZ^2\right) = \ZZ + s \ZZ$.

The excitation spectrum and gap locations of operators defined on $1D$ C\&P tilings are  also characterised by two integers  \eqref{eq:Bloch-relation}, a result quantified in the GLT~\citep{Bellissard_RMP_1991}.  
Winding numbers  $\winding_\phi\left[\Theta_s\right]$ in \eqref{eq:windings} at these gaps, are also  consequences of topological features of the tiling. 
As proven in \citep{Bellissard_RMP_1991} using the $K_0$ group, gap locations are obtained from another trace map here denoted $\tau_* ^{s} \left(K_0\right)$. In \citep{ADRS_2021_Bloch}, the equivalence of these two traces has been established for $1D$ C\&P quasiperiodic tilings, namely $\tau_* ^{d} \left(\smash{\CH}\right) = \tau_* ^{s} \left(K_0\right) = \ZZ + s \ZZ$, a result essentially expressed by \eqref{eq:Bloch-relation}.

However, as appealing as it may seem, this equivalence between diffraction and spectral features does not apply to all aperiodic tilings. The reason for that is to be found in the physical contents of the Čech cohomology group $\CH$ and the GLT. 
A main feature behind the GLT---sometimes hidden by the non trivial mathematics---is that the possible values $\IDOS$ of the IDOS on the gaps are given by the densities $\varrho_w$ of all possible words $w$ in the infinite tiling, viz.\ $\IDOS_w=\varrho_w$. Stated otherwise, for a gap to open in the IDOS, it is necessary that  some word appears infinitely many times (a condition also shared by periodic tilings). 
For the two-letter $\left\{ A,B\right\} $ alphabet, $A$ and $B$ tiles imply gaps on $\IDOS=\varrho_A$ and $\IDOS=\varrho_B$. To find the other densities for substitution tilings (see SM~\appref{app:sec:Gap-labeling}), one requires an updated rule $\sigma_n: \Gamma_n\to\Gamma_n$ with its occurrence matrix $M_n$ for each $n$. For primitive substitutions, the leading eigenvalue $\lambda_*$ is the same for all $M_n$, and the densities are given by the leading eigenvector $\boldsymbol{v}_{*,n}$. 
The eigenvectors $\boldsymbol{v}_*$ and $\boldsymbol{v}_{*,2}$ together with $\lambda_*$ suffice to span all other densities~\citep{Queffelec_Book_2010}. Thus, the GLT~\citep{Bellissard_RMP_1991} allows to compute
\begin{equation}
    \IDOS_{k,N}\left(E_g\right) = \frac{1}{a} \, \frac{k}{\lambda_*^N} \pmod{1},\quad k,N\in\NN,
    \label{eq:gap-labeling-formula}
\end{equation}
where $a$ is the least common multiplier of all elements of $\boldsymbol{v}_*$ and $\boldsymbol{v}_{*,2}$ (see SM~\appref{app:sec:Gap-labeling}). 
For quasiperiodic substitutions which have $\det(M)=\pm1$, \eqref{eq:gap-labeling-formula} reduces to
\begin{equation}
    \IDOS_{p,q}\left(E_g\right) = a^{-1} \left(p+q/\lambda_*\right) \pmod{1},\quad p,q\in\ZZ,
\end{equation}
namely, the densities for $w_n$ 
are integral linear combinations of $1$ and $s$ implying \eqref{eq:Bloch-relation} and  equivalent to $\tau_* ^{s} \left(K_0\right) = \ZZ + s \ZZ$. 
Therefore, the existence and location of gaps are essentially counting properties of tilings encapsulated in the distribution of word frequencies---a positive measure expressed by the normalised eigenvectors $\boldsymbol{v}_*$ and $\boldsymbol{v}_{*,2}$.  
In contrast, the cohomology group $\CH$ includes information on \emph{how to order} the words, i.e.\ their combinatorial properties; thus its relevance to the diffraction spectrum. 
But for $1D$ C\&P tilings, $\CH$ also accounts for words' densities $\varrho_w$, hence the equivalence $\tau_* ^{d} \left(\smash{\CH}\right) = \tau_* ^{s} \left(K_0\right) = \ZZ + s \ZZ$  \citep{ADRS_2021_Bloch} which generally holds for tilings with Bragg diffraction spectrum.


A remarkable example to sharpen this distinction is the Thue-Morse tiling: $\sigma_\text{TM}(A)=AB$, $\sigma_\text{TM}(B)=BA$.  It shares the same  occurrence matrix $M = \begin{psmallmatrix}1 & 1\\1 &1\end{psmallmatrix}$ with the periodic substitution $\sigma_\text{per}(A)=\sigma_\text{per}(B)=AB$.
Despite this, their structure factor and spectral gap distribution are completely different, a result to be associated to the different orderings of words of all lengths. From \eqref{eq:gap-labeling-formula}, the Thue-Morse tiling has a singular continuous distribution of gaps on $\IDOS=\frac13\,m/2^N$, but a complex and still unclear singular continuous diffraction component. 
Experimentally measured in \citep{Axel_1991}, diffraction peaks were announced to be located at $k/k_0=\frac13\,m/2^N$. A different formula $k/k_0=\frac{1}{p}\,m/2^N$ was reported by  \citep{Luck_PRB_1989,Cheng_1988,Wolny_2000a} ($p$ odd), \citep{Gazeau_2008} ($p$ prime), \citep{Kolar_1993} ($p=2^n-1$), and \citep{Baake_2014} (various $p$ with positive scaling exponents).
On top of that, Thue-Morse admits Bragg diffraction for $k_b/k_0=m,m+\frac12$ for $m\in\ZZ$~\citep{Kolar_1993}. This behaviour is not captured by the GLT formula  but it is implied by $\CH_{\text{TM}}\cong\ZZ\oplus\ZZ[\frac12]$ obtained from \eqref{eq:zeta}.
Hence, $\CH$ can detect complex structural behavior of tilings, but calculating the full diffraction spectrum may require further analysis \citep{ADRS_2021_Bloch}.
Another counterexample is provided by the Rudin-Shapiro tiling, which has continuous diffraction spectrum but a highly sparse  pure-point energy spectrum \citep{,Luck_PRB_1989,Bellissard_Review_1992,Luck_JPA_1993,Baake_Grimm_Book_2018,ADRS_2021_Bloch}.

Topology is often invoked to explain the resilience of specific properties to external perturbations, e.g.\ disorder. The underlying idea being that a physical quantity expressible as a topological invariant, will remain unchanged/protected against such perturbations, like the celebrated quantum Hall conductance directly related to a topological (Chern) number. It is interesting to assess this topological protection for aperiodic tilings. Topological quantities---either structural or spectral---result from 
$\CH$ and depend on sets of integers as in \eqref{eq:windings} and \eqref{eq:Bloch-relation} for winding numbers, Bragg peaks and spectral gaps locations. The sensitivity of these physical quantities has been tested against increasing disorder strength and indeed appeared to be surprisingly resilient \citep{Dareau_PRL_2017}.


\section*{Acknowledgments}

\begin{acknowledgments}
This work was supported by the Israel Science Foundation Grant No.~924/09 and the Pazy Research Foundation.
We thank C.~Schochet for fruitful discussions. 
\end{acknowledgments}

%


\begin{thebibliography}{48}%
\makeatletter
\providecommand \@ifxundefined [1]{%
 \@ifx{#1\undefined}
}%
\providecommand \@ifnum [1]{%
 \ifnum #1\expandafter \@firstoftwo
 \else \expandafter \@secondoftwo
 \fi
}%
\providecommand \@ifx [1]{%
 \ifx #1\expandafter \@firstoftwo
 \else \expandafter \@secondoftwo
 \fi
}%
\providecommand \natexlab [1]{#1}%
\providecommand \enquote  [1]{``#1''}%
\providecommand \bibnamefont  [1]{#1}%
\providecommand \bibfnamefont [1]{#1}%
\providecommand \citenamefont [1]{#1}%
\providecommand \href@noop [0]{\@secondoftwo}%
\providecommand \href [0]{\begingroup \@sanitize@url \@href}%
\providecommand \@href[1]{\@@startlink{#1}\@@href}%
\providecommand \@@href[1]{\endgroup#1\@@endlink}%
\providecommand \@sanitize@url [0]{\catcode `\\12\catcode `\$12\catcode
  `\&12\catcode `\#12\catcode `\^12\catcode `\_12\catcode `\%12\relax}%
\providecommand \@@startlink[1]{}%
\providecommand \@@endlink[0]{}%
\providecommand \url  [0]{\begingroup\@sanitize@url \@url }%
\providecommand \@url [1]{\endgroup\@href {#1}{\urlprefix }}%
\providecommand \urlprefix  [0]{URL }%
\providecommand \Eprint [0]{\href }%
\providecommand \doibase [0]{https://doi.org/}%
\providecommand \selectlanguage [0]{\@gobble}%
\providecommand \bibinfo  [0]{\@secondoftwo}%
\providecommand \bibfield  [0]{\@secondoftwo}%
\providecommand \translation [1]{[#1]}%
\providecommand \BibitemOpen [0]{}%
\providecommand \bibitemStop [0]{}%
\providecommand \bibitemNoStop [0]{.\EOS\space}%
\providecommand \EOS [0]{\spacefactor3000\relax}%
\providecommand \BibitemShut  [1]{\csname bibitem#1\endcsname}%
\let\auto@bib@innerbib\@empty
\bibitem [{\citenamefont {Senechal}(1996)}]{Senechal_Book_1996}%
  \BibitemOpen
  \bibfield  {author} {\bibinfo {author} {\bibfnamefont {M.}~\bibnamefont
  {Senechal}},\ }\href {https://books.google.co.il/books?id=LdQ8AAAAIAAJ}
  {\emph {\bibinfo {title} {{Quasicrystals and Geometry}}}}\ (\bibinfo
  {publisher} {Cambridge University Press},\ \bibinfo {year}
  {1996})\BibitemShut {NoStop}%
\bibitem [{\citenamefont {Mozes}(1989)}]{Mozes_JAM_1989}%
  \BibitemOpen
  \bibfield  {author} {\bibinfo {author} {\bibfnamefont {S.}~\bibnamefont
  {Mozes}},\ }\href {https://doi.org/10.1007/BF02793412} {\bibfield  {journal}
  {\bibinfo  {journal} {J. Anal. Math.}\ }\textbf {\bibinfo {volume} {53}},\
  \bibinfo {pages} {139} (\bibinfo {year} {1989})}\BibitemShut {NoStop}%
\bibitem [{\citenamefont {Levine}\ and\ \citenamefont
  {Steinhardt}(1984)}]{Levine_1984}%
  \BibitemOpen
  \bibfield  {author} {\bibinfo {author} {\bibfnamefont {D.}~\bibnamefont
  {Levine}}\ and\ \bibinfo {author} {\bibfnamefont {P.~J.}\ \bibnamefont
  {Steinhardt}},\ }\href {https://doi.org/10.1103/PhysRevLett.53.2477}
  {\bibfield  {journal} {\bibinfo  {journal} {Phys. Rev. Lett.}\ }\textbf
  {\bibinfo {volume} {53}},\ \bibinfo {pages} {2477} (\bibinfo {year}
  {1984})}\BibitemShut {NoStop}%
\bibitem [{\citenamefont {Shechtman}\ \emph {et~al.}(1984)\citenamefont
  {Shechtman}, \citenamefont {Blech}, \citenamefont {Gratias},\ and\
  \citenamefont {Cahn}}]{Shechtman_1984}%
  \BibitemOpen
  \bibfield  {author} {\bibinfo {author} {\bibfnamefont {D.}~\bibnamefont
  {Shechtman}}, \bibinfo {author} {\bibfnamefont {I.}~\bibnamefont {Blech}},
  \bibinfo {author} {\bibfnamefont {D.}~\bibnamefont {Gratias}},\ and\ \bibinfo
  {author} {\bibfnamefont {J.~W.}\ \bibnamefont {Cahn}},\ }\href
  {https://doi.org/10.1103/PhysRevLett.53.1951} {\bibfield  {journal} {\bibinfo
   {journal} {Phys. Rev. Lett.}\ }\textbf {\bibinfo {volume} {53}},\ \bibinfo
  {pages} {1951} (\bibinfo {year} {1984})}\BibitemShut {NoStop}%
\bibitem [{\citenamefont {Luck}(1989)}]{Luck_PRB_1989}%
  \BibitemOpen
  \bibfield  {author} {\bibinfo {author} {\bibfnamefont {J.~M.}\ \bibnamefont
  {Luck}},\ }\href {https://doi.org/10.1103/PhysRevB.39.5834} {\bibfield
  {journal} {\bibinfo  {journal} {Phys. Rev. B}\ }\textbf {\bibinfo {volume}
  {39}},\ \bibinfo {pages} {5834} (\bibinfo {year} {1989})}\BibitemShut
  {NoStop}%
\bibitem [{\citenamefont {Damanik}\ \emph {et~al.}(2008)\citenamefont
  {Damanik}, \citenamefont {Embree}, \citenamefont {Gorodetski},\ and\
  \citenamefont {Tcheremchantsev}}]{Damanik_CMP_2008}%
  \BibitemOpen
  \bibfield  {author} {\bibinfo {author} {\bibfnamefont {D.}~\bibnamefont
  {Damanik}}, \bibinfo {author} {\bibfnamefont {M.}~\bibnamefont {Embree}},
  \bibinfo {author} {\bibfnamefont {A.}~\bibnamefont {Gorodetski}},\ and\
  \bibinfo {author} {\bibfnamefont {S.}~\bibnamefont {Tcheremchantsev}},\
  }\href {https://doi.org/10.1007/s00220-008-0451-3} {\bibfield  {journal}
  {\bibinfo  {journal} {Commun. Math. Phys.}\ }\textbf {\bibinfo {volume}
  {280}},\ \bibinfo {pages} {499} (\bibinfo {year} {2008})}\BibitemShut
  {NoStop}%
\bibitem [{\citenamefont {Tanese}\ \emph {et~al.}(2014)\citenamefont {Tanese},
  \citenamefont {Gurevich}, \citenamefont {Baboux}, \citenamefont {Jacqmin},
  \citenamefont {Lema{\^{\i}}tre}, \citenamefont {Galopin}, \citenamefont
  {Sagnes}, \citenamefont {Amo}, \citenamefont {Bloch},\ and\ \citenamefont
  {Akkermans}}]{Tanese_2014}%
  \BibitemOpen
  \bibfield  {author} {\bibinfo {author} {\bibfnamefont {D.}~\bibnamefont
  {Tanese}}, \bibinfo {author} {\bibfnamefont {E.}~\bibnamefont {Gurevich}},
  \bibinfo {author} {\bibfnamefont {F.}~\bibnamefont {Baboux}}, \bibinfo
  {author} {\bibfnamefont {T.}~\bibnamefont {Jacqmin}}, \bibinfo {author}
  {\bibfnamefont {A.}~\bibnamefont {Lema{\^{\i}}tre}}, \bibinfo {author}
  {\bibfnamefont {E.}~\bibnamefont {Galopin}}, \bibinfo {author} {\bibfnamefont
  {I.}~\bibnamefont {Sagnes}}, \bibinfo {author} {\bibfnamefont
  {A.}~\bibnamefont {Amo}}, \bibinfo {author} {\bibfnamefont {J.}~\bibnamefont
  {Bloch}},\ and\ \bibinfo {author} {\bibfnamefont {E.}~\bibnamefont
  {Akkermans}},\ }\href {https://doi.org/10.1103/PhysRevLett.112.146404}
  {\bibfield  {journal} {\bibinfo  {journal} {Phys. Rev. Lett.}\ }\textbf
  {\bibinfo {volume} {112}},\ \bibinfo {pages} {146404} (\bibinfo {year}
  {2014})}\BibitemShut {NoStop}%
\bibitem [{\citenamefont {Sutherland}(1986)}]{Sutherland_PRB_1986}%
  \BibitemOpen
  \bibfield  {author} {\bibinfo {author} {\bibfnamefont {B.}~\bibnamefont
  {Sutherland}},\ }\href {https://doi.org/10.1103/PhysRevB.34.3904} {\bibfield
  {journal} {\bibinfo  {journal} {Phys. Rev. B}\ }\textbf {\bibinfo {volume}
  {34}},\ \bibinfo {pages} {3904} (\bibinfo {year} {1986})}\BibitemShut
  {NoStop}%
\bibitem [{\citenamefont {Bellissard}\ \emph {et~al.}(1992)\citenamefont
  {Bellissard}, \citenamefont {Bovier},\ and\ \citenamefont
  {Ghez}}]{Bellissard_RMP_1991}%
  \BibitemOpen
  \bibfield  {author} {\bibinfo {author} {\bibfnamefont {J.}~\bibnamefont
  {Bellissard}}, \bibinfo {author} {\bibfnamefont {A.}~\bibnamefont {Bovier}},\
  and\ \bibinfo {author} {\bibfnamefont {J.-M.}\ \bibnamefont {Ghez}},\ }\href
  {https://doi.org/10.1142/S0129055X92000029} {\bibfield  {journal} {\bibinfo
  {journal} {Rev. Math. Phys.}\ }\textbf {\bibinfo {volume} {04}},\ \bibinfo
  {pages} {1} (\bibinfo {year} {1992})}\BibitemShut {NoStop}%
\bibitem [{\citenamefont {Bellissard}(1992)}]{Bellissard_Review_1992}%
  \BibitemOpen
  \bibfield  {author} {\bibinfo {author} {\bibfnamefont {J.}~\bibnamefont
  {Bellissard}},\ }in\ \href {https://doi.org/10.1007/978-3-662-02838-4_12}
  {\emph {\bibinfo {booktitle} {From Number Theory to Physics}}},\ \bibinfo
  {series and number} {Les Houches March '89},\ \bibinfo {editor} {edited by\
  \bibinfo {editor} {\bibfnamefont {M.}~\bibnamefont {Waldschmidt}}, \bibinfo
  {editor} {\bibfnamefont {P.}~\bibnamefont {Moussa}}, \bibinfo {editor}
  {\bibfnamefont {J.-M.}\ \bibnamefont {Luck}},\ and\ \bibinfo {editor}
  {\bibfnamefont {C.}~\bibnamefont {Itzykson}}}\ (\bibinfo  {publisher}
  {Springer Berlin Heidelberg},\ \bibinfo {year} {1992})\ pp.\ \bibinfo {pages}
  {538--630}\BibitemShut {NoStop}%
\bibitem [{\citenamefont {Thouless}\ \emph {et~al.}(1982)\citenamefont
  {Thouless}, \citenamefont {Kohmoto}, \citenamefont {Nightingale},\ and\
  \citenamefont {den Nijs}}]{TKNN_1982}%
  \BibitemOpen
  \bibfield  {author} {\bibinfo {author} {\bibfnamefont {D.~J.}\ \bibnamefont
  {Thouless}}, \bibinfo {author} {\bibfnamefont {M.}~\bibnamefont {Kohmoto}},
  \bibinfo {author} {\bibfnamefont {M.~P.}\ \bibnamefont {Nightingale}},\ and\
  \bibinfo {author} {\bibfnamefont {M.}~\bibnamefont {den Nijs}},\ }\href
  {https://doi.org/10.1103/PhysRevLett.49.405} {\bibfield  {journal} {\bibinfo
  {journal} {Phys. Rev. Lett.}\ }\textbf {\bibinfo {volume} {49}},\ \bibinfo
  {pages} {405} (\bibinfo {year} {1982})}\BibitemShut {NoStop}%
\bibitem [{\citenamefont {Berry}(1984)}]{Berry_1984}%
  \BibitemOpen
  \bibfield  {author} {\bibinfo {author} {\bibfnamefont {M.~V.}\ \bibnamefont
  {Berry}},\ }\href {https://doi.org/10.1098/rspa.1984.0023} {\bibfield
  {journal} {\bibinfo  {journal} {Proc. R. Soc. Lond. A}\ }\textbf {\bibinfo
  {volume} {392}},\ \bibinfo {pages} {45} (\bibinfo {year} {1984})}\BibitemShut
  {NoStop}%
\bibitem [{\citenamefont {Thouless}(1983)}]{Thouless_1983}%
  \BibitemOpen
  \bibfield  {author} {\bibinfo {author} {\bibfnamefont {D.~J.}\ \bibnamefont
  {Thouless}},\ }\href {https://doi.org/10.1103/physrevb.27.6083} {\bibfield
  {journal} {\bibinfo  {journal} {Phys. Rev. B}\ }\textbf {\bibinfo {volume}
  {27}},\ \bibinfo {pages} {6083} (\bibinfo {year} {1983})}\BibitemShut
  {NoStop}%
\bibitem [{\citenamefont {Xiao}\ \emph {et~al.}(2010)\citenamefont {Xiao},
  \citenamefont {Chang},\ and\ \citenamefont {Niu}}]{Xiao_2010}%
  \BibitemOpen
  \bibfield  {author} {\bibinfo {author} {\bibfnamefont {D.}~\bibnamefont
  {Xiao}}, \bibinfo {author} {\bibfnamefont {M.-C.}\ \bibnamefont {Chang}},\
  and\ \bibinfo {author} {\bibfnamefont {Q.}~\bibnamefont {Niu}},\ }\href
  {https://doi.org/10.1103/revmodphys.82.1959} {\bibfield  {journal} {\bibinfo
  {journal} {Rev. Mod. Phys.}\ }\textbf {\bibinfo {volume} {82}},\ \bibinfo
  {pages} {1959} (\bibinfo {year} {2010})}\BibitemShut {NoStop}%
\bibitem [{\citenamefont {Kunz}(1986)}]{Kunz_1986}%
  \BibitemOpen
  \bibfield  {author} {\bibinfo {author} {\bibfnamefont {H.}~\bibnamefont
  {Kunz}},\ }\href {https://doi.org/10.1103/PhysRevLett.57.1095} {\bibfield
  {journal} {\bibinfo  {journal} {Phys. Rev. Lett.}\ }\textbf {\bibinfo
  {volume} {57}},\ \bibinfo {pages} {1095} (\bibinfo {year}
  {1986})}\BibitemShut {NoStop}%
\bibitem [{\citenamefont {Poddubny}\ and\ \citenamefont
  {Ivchenko}(2010)}]{Poddubny_Ivchenko_PE_2010}%
  \BibitemOpen
  \bibfield  {author} {\bibinfo {author} {\bibfnamefont {A.}~\bibnamefont
  {Poddubny}}\ and\ \bibinfo {author} {\bibfnamefont {E.}~\bibnamefont
  {Ivchenko}},\ }\href
  {https://doi.org/http://dx.doi.org/10.1016/j.physe.2010.02.020} {\bibfield
  {journal} {\bibinfo  {journal} {Physica E}\ }\textbf {\bibinfo {volume}
  {42}},\ \bibinfo {pages} {1871} (\bibinfo {year} {2010})}\BibitemShut
  {NoStop}%
\bibitem [{\citenamefont {Kraus}\ \emph {et~al.}(2012)\citenamefont {Kraus},
  \citenamefont {Lahini}, \citenamefont {Ringel}, \citenamefont {Verbin},\ and\
  \citenamefont {Zilberberg}}]{Kraus_2012a}%
  \BibitemOpen
  \bibfield  {author} {\bibinfo {author} {\bibfnamefont {Y.~E.}\ \bibnamefont
  {Kraus}}, \bibinfo {author} {\bibfnamefont {Y.}~\bibnamefont {Lahini}},
  \bibinfo {author} {\bibfnamefont {Z.}~\bibnamefont {Ringel}}, \bibinfo
  {author} {\bibfnamefont {M.}~\bibnamefont {Verbin}},\ and\ \bibinfo {author}
  {\bibfnamefont {O.}~\bibnamefont {Zilberberg}},\ }\href
  {https://doi.org/10.1103/PhysRevLett.109.106402} {\bibfield  {journal}
  {\bibinfo  {journal} {Phys. Rev. Lett.}\ }\textbf {\bibinfo {volume} {109}},\
  \bibinfo {pages} {106402} (\bibinfo {year} {2012})}\BibitemShut {NoStop}%
\bibitem [{\citenamefont {Kraus}\ and\ \citenamefont
  {Zilberberg}(2012)}]{Kraus_2012b}%
  \BibitemOpen
  \bibfield  {author} {\bibinfo {author} {\bibfnamefont {Y.~E.}\ \bibnamefont
  {Kraus}}\ and\ \bibinfo {author} {\bibfnamefont {O.}~\bibnamefont
  {Zilberberg}},\ }\href {https://doi.org/10.1103/PhysRevLett.109.116404}
  {\bibfield  {journal} {\bibinfo  {journal} {Phys. Rev. Lett.}\ }\textbf
  {\bibinfo {volume} {109}},\ \bibinfo {pages} {116404} (\bibinfo {year}
  {2012})}\BibitemShut {NoStop}%
\bibitem [{\citenamefont {Bandres}\ \emph {et~al.}(2016)\citenamefont
  {Bandres}, \citenamefont {Rechtsman},\ and\ \citenamefont
  {Segev}}]{Bandres_2016}%
  \BibitemOpen
  \bibfield  {author} {\bibinfo {author} {\bibfnamefont {M.~A.}\ \bibnamefont
  {Bandres}}, \bibinfo {author} {\bibfnamefont {M.~C.}\ \bibnamefont
  {Rechtsman}},\ and\ \bibinfo {author} {\bibfnamefont {M.}~\bibnamefont
  {Segev}},\ }\href {https://doi.org/10.1103/PhysRevX.6.011016} {\bibfield
  {journal} {\bibinfo  {journal} {Phys. Rev. X}\ }\textbf {\bibinfo {volume}
  {6}},\ \bibinfo {pages} {011016} (\bibinfo {year} {2016})}\BibitemShut
  {NoStop}%
\bibitem [{\citenamefont {Baboux}\ \emph {et~al.}(2017)\citenamefont {Baboux},
  \citenamefont {Levy}, \citenamefont {Lema{\^{\i}}tre}, \citenamefont
  {G{\'{o}}mez}, \citenamefont {Galopin}, \citenamefont {Gratiet},
  \citenamefont {Sagnes}, \citenamefont {Amo}, \citenamefont {Bloch},\ and\
  \citenamefont {Akkermans}}]{Baboux_2017}%
  \BibitemOpen
  \bibfield  {author} {\bibinfo {author} {\bibfnamefont {F.}~\bibnamefont
  {Baboux}}, \bibinfo {author} {\bibfnamefont {E.}~\bibnamefont {Levy}},
  \bibinfo {author} {\bibfnamefont {A.}~\bibnamefont {Lema{\^{\i}}tre}},
  \bibinfo {author} {\bibfnamefont {C.}~\bibnamefont {G{\'{o}}mez}}, \bibinfo
  {author} {\bibfnamefont {E.}~\bibnamefont {Galopin}}, \bibinfo {author}
  {\bibfnamefont {L.~L.}\ \bibnamefont {Gratiet}}, \bibinfo {author}
  {\bibfnamefont {I.}~\bibnamefont {Sagnes}}, \bibinfo {author} {\bibfnamefont
  {A.}~\bibnamefont {Amo}}, \bibinfo {author} {\bibfnamefont {J.}~\bibnamefont
  {Bloch}},\ and\ \bibinfo {author} {\bibfnamefont {E.}~\bibnamefont
  {Akkermans}},\ }\href {https://doi.org/10.1103/PhysRevB.95.161114} {\bibfield
   {journal} {\bibinfo  {journal} {Phys. Rev. B}\ }\textbf {\bibinfo {volume}
  {95}},\ \bibinfo {pages} {161114} (\bibinfo {year} {2017})}\BibitemShut
  {NoStop}%
\bibitem [{\citenamefont {Dareau}\ \emph {et~al.}(2017)\citenamefont {Dareau},
  \citenamefont {Levy}, \citenamefont {Aguilera}, \citenamefont {Bouganne},
  \citenamefont {Akkermans}, \citenamefont {Gerbier},\ and\ \citenamefont
  {Beugnon}}]{Dareau_PRL_2017}%
  \BibitemOpen
  \bibfield  {author} {\bibinfo {author} {\bibfnamefont {A.}~\bibnamefont
  {Dareau}}, \bibinfo {author} {\bibfnamefont {E.}~\bibnamefont {Levy}},
  \bibinfo {author} {\bibfnamefont {M.~B.}\ \bibnamefont {Aguilera}}, \bibinfo
  {author} {\bibfnamefont {R.}~\bibnamefont {Bouganne}}, \bibinfo {author}
  {\bibfnamefont {E.}~\bibnamefont {Akkermans}}, \bibinfo {author}
  {\bibfnamefont {F.}~\bibnamefont {Gerbier}},\ and\ \bibinfo {author}
  {\bibfnamefont {J.}~\bibnamefont {Beugnon}},\ }\href
  {https://doi.org/10.1103/PhysRevLett.119.215304} {\bibfield  {journal}
  {\bibinfo  {journal} {Phys. Rev. Lett.}\ }\textbf {\bibinfo {volume} {119}},\
  \bibinfo {pages} {215304} (\bibinfo {year} {2017})}\BibitemShut {NoStop}%
\bibitem [{\citenamefont {Dana}(2014)}]{Dana_2014}%
  \BibitemOpen
  \bibfield  {author} {\bibinfo {author} {\bibfnamefont {I.}~\bibnamefont
  {Dana}},\ }\href {https://doi.org/10.1103/PhysRevB.89.205111} {\bibfield
  {journal} {\bibinfo  {journal} {Phys. Rev. B}\ }\textbf {\bibinfo {volume}
  {89}},\ \bibinfo {pages} {205111} (\bibinfo {year} {2014})}\BibitemShut
  {NoStop}%
\bibitem [{\citenamefont {Poshakinskiy}\ \emph {et~al.}(2015)\citenamefont
  {Poshakinskiy}, \citenamefont {Poddubny},\ and\ \citenamefont
  {Hafezi}}]{Poshakinskiy_2015}%
  \BibitemOpen
  \bibfield  {author} {\bibinfo {author} {\bibfnamefont {A.~V.}\ \bibnamefont
  {Poshakinskiy}}, \bibinfo {author} {\bibfnamefont {A.~N.}\ \bibnamefont
  {Poddubny}},\ and\ \bibinfo {author} {\bibfnamefont {M.}~\bibnamefont
  {Hafezi}},\ }\href {https://doi.org/10.1103/PhysRevA.91.043830} {\bibfield
  {journal} {\bibinfo  {journal} {Phys. Rev. A}\ }\textbf {\bibinfo {volume}
  {91}},\ \bibinfo {pages} {043830} (\bibinfo {year} {2015})}\BibitemShut
  {NoStop}%
\bibitem [{\citenamefont {Madsen}\ \emph {et~al.}(2013)\citenamefont {Madsen},
  \citenamefont {Bergholtz},\ and\ \citenamefont {Brouwer}}]{Madsen_2013}%
  \BibitemOpen
  \bibfield  {author} {\bibinfo {author} {\bibfnamefont {K.~A.}\ \bibnamefont
  {Madsen}}, \bibinfo {author} {\bibfnamefont {E.~J.}\ \bibnamefont
  {Bergholtz}},\ and\ \bibinfo {author} {\bibfnamefont {P.~W.}\ \bibnamefont
  {Brouwer}},\ }\href {https://doi.org/10.1103/PhysRevB.88.125118} {\bibfield
  {journal} {\bibinfo  {journal} {Phys. Rev. B}\ }\textbf {\bibinfo {volume}
  {88}},\ \bibinfo {pages} {125118} (\bibinfo {year} {2013})}\BibitemShut
  {NoStop}%
\bibitem [{\citenamefont {Akkermans}\ \emph {et~al.}(2021)\citenamefont
  {Akkermans}, \citenamefont {Don}, \citenamefont {Rosenberg},\ and\
  \citenamefont {Schochet}}]{ADRS_2021_Bloch}%
  \BibitemOpen
  \bibfield  {author} {\bibinfo {author} {\bibfnamefont {E.}~\bibnamefont
  {Akkermans}}, \bibinfo {author} {\bibfnamefont {Y.}~\bibnamefont {Don}},
  \bibinfo {author} {\bibfnamefont {J.}~\bibnamefont {Rosenberg}},\ and\
  \bibinfo {author} {\bibfnamefont {C.~L.}\ \bibnamefont {Schochet}},\ }\href
  {https://doi.org/https://doi.org/10.1016/j.geomphys.2021.104217} {\bibfield
  {journal} {\bibinfo  {journal} {J. Geom. Phys.}\ }\textbf {\bibinfo {volume}
  {165}},\ \bibinfo {pages} {104217} (\bibinfo {year} {2021})}\BibitemShut
  {NoStop}%
\bibitem [{\citenamefont {Akkermans}\ \emph {et~al.}(2013)\citenamefont
  {Akkermans}, \citenamefont {Dunne},\ and\ \citenamefont
  {Levy}}]{Akkermans_Review_2014}%
  \BibitemOpen
  \bibfield  {author} {\bibinfo {author} {\bibfnamefont {E.}~\bibnamefont
  {Akkermans}}, \bibinfo {author} {\bibfnamefont {G.~V.}\ \bibnamefont
  {Dunne}},\ and\ \bibinfo {author} {\bibfnamefont {E.}~\bibnamefont {Levy}},\
  }in\ \href {https://doi.org/10.1201/b15653-13} {\emph {\bibinfo {booktitle}
  {Optics of Aperiodic Structures: Fundamentals and Device Applications}}},\
  \bibinfo {editor} {edited by\ \bibinfo {editor} {\bibfnamefont
  {L.}~\bibnamefont {{Dal Negro}}}}\ (\bibinfo  {publisher} {Pan Stanford
  Publishing},\ \bibinfo {year} {2013})\ pp.\ \bibinfo {pages}
  {407--449}\BibitemShut {NoStop}%
\bibitem [{\citenamefont {Don}\ and\ \citenamefont
  {Akkermans}(2021)}]{Supplementary}%
  \BibitemOpen
  \bibfield  {author} {\bibinfo {author} {\bibfnamefont {Y.}~\bibnamefont
  {Don}}\ and\ \bibinfo {author} {\bibfnamefont {E.}~\bibnamefont
  {Akkermans}},\ }\href@noop {} {} (\bibinfo {year} {2021}),\ \bibinfo {note}
  {{See Supplementary Material}}\BibitemShut {NoStop}%
\bibitem [{\citenamefont {Levy}\ \emph {et~al.}(2015)\citenamefont {Levy},
  \citenamefont {Barak}, \citenamefont {Fisher},\ and\ \citenamefont
  {Akkermans}}]{Levy_arXiv_2015}%
  \BibitemOpen
  \bibfield  {author} {\bibinfo {author} {\bibfnamefont {E.}~\bibnamefont
  {Levy}}, \bibinfo {author} {\bibfnamefont {A.}~\bibnamefont {Barak}},
  \bibinfo {author} {\bibfnamefont {A.}~\bibnamefont {Fisher}},\ and\ \bibinfo
  {author} {\bibfnamefont {E.}~\bibnamefont {Akkermans}},\ }\href@noop {}
  {\bibinfo {title} {{Topological properties of Fibonacci quasicrystals : A
  scattering analysis of Chern numbers}}} (\bibinfo {year} {2015}),\ \Eprint
  {https://arxiv.org/abs/1509.04028} {arXiv:1509.04028 [physics.optics]}
  \BibitemShut {NoStop}%
\bibitem [{\citenamefont {Levy}\ and\ \citenamefont
  {Akkermans}(2017)}]{Levy_EPJ_2017}%
  \BibitemOpen
  \bibfield  {author} {\bibinfo {author} {\bibfnamefont {E.}~\bibnamefont
  {Levy}}\ and\ \bibinfo {author} {\bibfnamefont {E.}~\bibnamefont
  {Akkermans}},\ }\href {https://doi.org/10.1140/epjst/e2016-60341-8}
  {\bibfield  {journal} {\bibinfo  {journal} {Eur. Phys. J. Special Topics}\
  }\textbf {\bibinfo {volume} {226}},\ \bibinfo {pages} {1563} (\bibinfo {year}
  {2017})}\BibitemShut {NoStop}%
\bibitem [{\citenamefont {Luck}(1994)}]{Luck_Review_1994}%
  \BibitemOpen
  \bibfield  {author} {\bibinfo {author} {\bibfnamefont {J.~M.}\ \bibnamefont
  {Luck}},\ }in\ \href
  {https://doi.org/http://dx.doi.org/10.1016/B978-0-444-81591-0.50010-9} {\emph
  {\bibinfo {booktitle} {{Fundamental Problems in Statistical Mechanics,
  VIII}}}},\ \bibinfo {editor} {edited by\ \bibinfo {editor} {\bibfnamefont
  {M.~H.}\ \bibnamefont {Ernst}}\ and\ \bibinfo {editor} {\bibfnamefont
  {H.}~\bibnamefont {{van Beijeren}}}}\ (\bibinfo  {publisher} {Elsevier},\
  \bibinfo {year} {1994})\ pp.\ \bibinfo {pages} {127--167}\BibitemShut
  {NoStop}%
\bibitem [{\citenamefont {Duneau}\ and\ \citenamefont
  {Katz}(1985)}]{DuneauKatz_1985}%
  \BibitemOpen
  \bibfield  {author} {\bibinfo {author} {\bibfnamefont {M.}~\bibnamefont
  {Duneau}}\ and\ \bibinfo {author} {\bibfnamefont {A.}~\bibnamefont {Katz}},\
  }\href {https://doi.org/10.1103/PhysRevLett.54.2688} {\bibfield  {journal}
  {\bibinfo  {journal} {Phys. Rev. Lett.}\ }\textbf {\bibinfo {volume} {54}},\
  \bibinfo {pages} {2688} (\bibinfo {year} {1985})}\BibitemShut {NoStop}%
\bibitem [{\citenamefont {Katz}\ and\ \citenamefont
  {Duneau}(1986)}]{KatzDuneau_1986}%
  \BibitemOpen
  \bibfield  {author} {\bibinfo {author} {\bibfnamefont {A.}~\bibnamefont
  {Katz}}\ and\ \bibinfo {author} {\bibfnamefont {M.}~\bibnamefont {Duneau}},\
  }\href {https://doi.org/10.1051/jphys:01986004702018100} {\bibfield
  {journal} {\bibinfo  {journal} {J. Phys. France}\ }\textbf {\bibinfo {volume}
  {47}},\ \bibinfo {pages} {181} (\bibinfo {year} {1986})}\BibitemShut
  {NoStop}%
\bibitem [{\citenamefont {Julien}(2009)}]{Julien_2009}%
  \BibitemOpen
  \bibfield  {author} {\bibinfo {author} {\bibfnamefont {A.}~\bibnamefont
  {Julien}},\ }\href {https://doi.org/10.1017/s0143385709000194} {\bibfield
  {journal} {\bibinfo  {journal} {Ergod. Th. Dynam. Sys.}\ }\textbf {\bibinfo
  {volume} {30}},\ \bibinfo {pages} {489} (\bibinfo {year} {2009})}\BibitemShut
  {NoStop}%
\bibitem [{\citenamefont {Arthur~Robinson}(2004)}]{Robinson_Review_2004}%
  \BibitemOpen
  \bibfield  {author} {\bibinfo {author} {\bibfnamefont {E.}~\bibnamefont
  {Arthur~Robinson}, \bibfnamefont {Jr.}},\ }in\ \href
  {https://doi.org/10.1090/psapm/060} {\emph {\bibinfo {booktitle} {Symbolic
  dynamics and its applications}}},\ \bibinfo {series} {Proceedings of Symposia
  in Applied Mathematics}, Vol.~\bibinfo {volume} {60},\ \bibinfo {editor}
  {edited by\ \bibinfo {editor} {\bibfnamefont {S.}~\bibnamefont {Williams}}}\
  (\bibinfo  {publisher} {American Mathematical Society},\ \bibinfo {year}
  {2004})\ pp.\ \bibinfo {pages} {81--120}\BibitemShut {NoStop}%
\bibitem [{\citenamefont {Don}(2021)}]{Don_Thesis_2021}%
  \BibitemOpen
  \bibfield  {author} {\bibinfo {author} {\bibfnamefont {Y.}~\bibnamefont
  {Don}},\ }\emph {\bibinfo {title} {{Topological Properties of Aperiodic
  Tilings and Fractals}}},\ \href
  {http://www.graduate.technion.ac.il/Theses/Abstracts.asp?Id=31197} {\bibinfo
  {type} {{PhD Thesis}}},\ \bibinfo  {school} {Technion -- Israel Institute of
  Technology} (\bibinfo {year} {2021})\BibitemShut {NoStop}%
\bibitem [{\citenamefont {Johnson}\ and\ \citenamefont
  {Moser}(1982)}]{Johnson_1982}%
  \BibitemOpen
  \bibfield  {author} {\bibinfo {author} {\bibfnamefont {R.}~\bibnamefont
  {Johnson}}\ and\ \bibinfo {author} {\bibfnamefont {J.}~\bibnamefont
  {Moser}},\ }\href {https://doi.org/10.1007/bf01208484} {\bibfield  {journal}
  {\bibinfo  {journal} {Commun. Math. Phys.}\ }\textbf {\bibinfo {volume}
  {84}},\ \bibinfo {pages} {403} (\bibinfo {year} {1982})}\BibitemShut
  {NoStop}%
\bibitem [{\citenamefont {Johnson}(1986)}]{Johnson_1986}%
  \BibitemOpen
  \bibfield  {author} {\bibinfo {author} {\bibfnamefont {R.~A.}\ \bibnamefont
  {Johnson}},\ }\href {https://doi.org/10.1016/0022-0396(86)90125-7} {\bibfield
   {journal} {\bibinfo  {journal} {J. Differ. Equ.}\ }\textbf {\bibinfo
  {volume} {61}},\ \bibinfo {pages} {54} (\bibinfo {year} {1986})}\BibitemShut
  {NoStop}%
\bibitem [{\citenamefont {Delyon}\ and\ \citenamefont
  {Souillard}(1983)}]{Delyon_1983}%
  \BibitemOpen
  \bibfield  {author} {\bibinfo {author} {\bibfnamefont {F.}~\bibnamefont
  {Delyon}}\ and\ \bibinfo {author} {\bibfnamefont {B.}~\bibnamefont
  {Souillard}},\ }\href {https://projecteuclid.org/euclid.cmp/1103922817}
  {\bibfield  {journal} {\bibinfo  {journal} {Commun. Math. Phys.}\ }\textbf
  {\bibinfo {volume} {89}},\ \bibinfo {pages} {415} (\bibinfo {year}
  {1983})}\BibitemShut {NoStop}%
\bibitem [{\citenamefont {Luck}\ \emph {et~al.}(1993)\citenamefont {Luck},
  \citenamefont {Godr\`{e}che}, \citenamefont {Janner},\ and\ \citenamefont
  {Janssen}}]{Luck_JPA_1993}%
  \BibitemOpen
  \bibfield  {author} {\bibinfo {author} {\bibfnamefont {J.~M.}\ \bibnamefont
  {Luck}}, \bibinfo {author} {\bibfnamefont {C.}~\bibnamefont {Godr\`{e}che}},
  \bibinfo {author} {\bibfnamefont {A.}~\bibnamefont {Janner}},\ and\ \bibinfo
  {author} {\bibfnamefont {T.}~\bibnamefont {Janssen}},\ }\href
  {http://stacks.iop.org/0305-4470/26/i=8/a=020} {\bibfield  {journal}
  {\bibinfo  {journal} {J. Phys. A}\ }\textbf {\bibinfo {volume} {26}},\
  \bibinfo {pages} {1951} (\bibinfo {year} {1993})}\BibitemShut {NoStop}%
\bibitem [{\citenamefont {Baake}\ and\ \citenamefont
  {Grimm}(2018)}]{Baake_Grimm_Book_2018}%
  \BibitemOpen
  \bibfield  {author} {\bibinfo {author} {\bibfnamefont {M.}~\bibnamefont
  {Baake}}\ and\ \bibinfo {author} {\bibfnamefont {U.}~\bibnamefont {Grimm}},\
  }\href {https://www.cambridge.org/9780521869928} {\emph {\bibinfo {title}
  {{Aperiodic Order}}}},\ \bibinfo {series} {Encyclopedia of Mathematics and
  its Applications}, Vol.\ \bibinfo {volume} {2: Crystallography and Almost
  Periodicity}\ (\bibinfo  {publisher} {Cambridge University Press},\ \bibinfo
  {year} {2018})\BibitemShut {NoStop}%
\bibitem [{\citenamefont {Anderson}\ and\ \citenamefont
  {Putnam}(1998)}]{Anderson_Putnam_ETDS_1998}%
  \BibitemOpen
  \bibfield  {author} {\bibinfo {author} {\bibfnamefont {J.~E.}\ \bibnamefont
  {Anderson}}\ and\ \bibinfo {author} {\bibfnamefont {I.~F.}\ \bibnamefont
  {Putnam}},\ }\href
  {https://www.cambridge.org/core/journals/ergodic-theory-and-dynamical-systems/article/div-classtitletopological-invariants-for-substitution-tilings-and-their-associated-cast-algebrasdiv/5D7DB90543165287CA5C4CBF012323F5}
  {\bibfield  {journal} {\bibinfo  {journal} {Ergod. Th. Dynam. Sys.}\ }\textbf
  {\bibinfo {volume} {18}},\ \bibinfo {pages} {509} (\bibinfo {year}
  {1998})}\BibitemShut {NoStop}%
\bibitem [{\citenamefont {Queff\'elec}(2010)}]{Queffelec_Book_2010}%
  \BibitemOpen
  \bibfield  {author} {\bibinfo {author} {\bibfnamefont {M.}~\bibnamefont
  {Queff\'elec}},\ }\href
  {http://link.springer.com/book/10.1007\%2F978-3-642-11212-6} {\emph {\bibinfo
  {title} {{Substitution Dynamical Systems -- Spectral Analysis}}}},\ \bibinfo
  {edition} {2nd}\ ed.,\ \bibinfo {series} {Lecture Notes in Mathematics},
  Vol.\ \bibinfo {volume} {1294}\ (\bibinfo  {publisher} {Springer Berlin
  Heidelberg},\ \bibinfo {year} {2010})\BibitemShut {NoStop}%
\bibitem [{\citenamefont {Axel}\ and\ \citenamefont
  {Terauchi}(1991)}]{Axel_1991}%
  \BibitemOpen
  \bibfield  {author} {\bibinfo {author} {\bibfnamefont {F.}~\bibnamefont
  {Axel}}\ and\ \bibinfo {author} {\bibfnamefont {H.}~\bibnamefont
  {Terauchi}},\ }\href {https://doi.org/10.1103/PhysRevLett.66.2223} {\bibfield
   {journal} {\bibinfo  {journal} {Phys. Rev. Lett.}\ }\textbf {\bibinfo
  {volume} {66}},\ \bibinfo {pages} {2223} (\bibinfo {year}
  {1991})}\BibitemShut {NoStop}%
\bibitem [{\citenamefont {Cheng}\ \emph {et~al.}(1988)\citenamefont {Cheng},
  \citenamefont {Savit},\ and\ \citenamefont {Merlin}}]{Cheng_1988}%
  \BibitemOpen
  \bibfield  {author} {\bibinfo {author} {\bibfnamefont {Z.}~\bibnamefont
  {Cheng}}, \bibinfo {author} {\bibfnamefont {R.}~\bibnamefont {Savit}},\ and\
  \bibinfo {author} {\bibfnamefont {R.}~\bibnamefont {Merlin}},\ }\href
  {https://doi.org/10.1103/PhysRevB.37.4375} {\bibfield  {journal} {\bibinfo
  {journal} {Phys. Rev. B}\ }\textbf {\bibinfo {volume} {37}},\ \bibinfo
  {pages} {4375} (\bibinfo {year} {1988})}\BibitemShut {NoStop}%
\bibitem [{\citenamefont {Wolny}\ \emph {et~al.}(2000)\citenamefont {Wolny},
  \citenamefont {Wn{\k{e}}k}, \citenamefont {Verger-Gaugry},\ and\
  \citenamefont {Pytlik}}]{Wolny_2000a}%
  \BibitemOpen
  \bibfield  {author} {\bibinfo {author} {\bibfnamefont {J.}~\bibnamefont
  {Wolny}}, \bibinfo {author} {\bibfnamefont {A.}~\bibnamefont {Wn{\k{e}}k}},
  \bibinfo {author} {\bibfnamefont {J.-L.}\ \bibnamefont {Verger-Gaugry}},\
  and\ \bibinfo {author} {\bibfnamefont {L.}~\bibnamefont {Pytlik}},\ }\href
  {https://doi.org/10.1016/s0921-5093(00)01154-0} {\bibfield  {journal}
  {\bibinfo  {journal} {Mater. Sci. Eng. A}\ }\textbf {\bibinfo {volume}
  {294-296}},\ \bibinfo {pages} {381} (\bibinfo {year} {2000})}\BibitemShut
  {NoStop}%
\bibitem [{\citenamefont {Gazeau}\ and\ \citenamefont
  {Verger-Gaugry}(2008)}]{Gazeau_2008}%
  \BibitemOpen
  \bibfield  {author} {\bibinfo {author} {\bibfnamefont {J.-P.}\ \bibnamefont
  {Gazeau}}\ and\ \bibinfo {author} {\bibfnamefont {J.-L.}\ \bibnamefont
  {Verger-Gaugry}},\ }\href {https://doi.org/10.5802/jtnb.645} {\bibfield
  {journal} {\bibinfo  {journal} {J. Th\'{e}or. Nr. Bordx.}\ }\textbf {\bibinfo
  {volume} {20}},\ \bibinfo {pages} {673} (\bibinfo {year} {2008})}\BibitemShut
  {NoStop}%
\bibitem [{\citenamefont {Kol\'{a}\v{r}}\ \emph {et~al.}(1993)\citenamefont
  {Kol\'{a}\v{r}}, \citenamefont {Iochum},\ and\ \citenamefont
  {Raymond}}]{Kolar_1993}%
  \BibitemOpen
  \bibfield  {author} {\bibinfo {author} {\bibfnamefont {M.}~\bibnamefont
  {Kol\'{a}\v{r}}}, \bibinfo {author} {\bibfnamefont {B.}~\bibnamefont
  {Iochum}},\ and\ \bibinfo {author} {\bibfnamefont {L.}~\bibnamefont
  {Raymond}},\ }\href {https://doi.org/10.1088/0305-4470/26/24/011} {\bibfield
  {journal} {\bibinfo  {journal} {J. Phys. A: Math. Gen.}\ }\textbf {\bibinfo
  {volume} {26}},\ \bibinfo {pages} {7343} (\bibinfo {year}
  {1993})}\BibitemShut {NoStop}%
\bibitem [{\citenamefont {Baake}\ \emph {et~al.}(2014)\citenamefont {Baake},
  \citenamefont {Grimm},\ and\ \citenamefont {Nilsson}}]{Baake_2014}%
  \BibitemOpen
  \bibfield  {author} {\bibinfo {author} {\bibfnamefont {M.}~\bibnamefont
  {Baake}}, \bibinfo {author} {\bibfnamefont {U.}~\bibnamefont {Grimm}},\ and\
  \bibinfo {author} {\bibfnamefont {J.}~\bibnamefont {Nilsson}},\ }\href
  {https://doi.org/10.12693/aphyspola.126.431} {\bibfield  {journal} {\bibinfo
  {journal} {Acta Phys. Pol. A}\ }\textbf {\bibinfo {volume} {126}},\ \bibinfo
  {pages} {431} (\bibinfo {year} {2014})}\BibitemShut {NoStop}%
\end{thebibliography}


\onecolumngrid
\clearpage

\supplementary

\end{document}